\documentclass[journal]{IEEEtran}
\usepackage{bbm}
\usepackage{caption}
\usepackage{subfigure}
\usepackage{graphicx}
\usepackage{latexsym}
\usepackage{diagbox}
\usepackage{changepage}
\usepackage[fleqn]{amsmath}
\usepackage{amsmath}
\usepackage{amsfonts}
\usepackage{indentfirst}
\usepackage{CJK}
\usepackage{indentfirst}
\usepackage[varg]{txfonts}
\usepackage{stfloats}
\usepackage{multirow}%
\usepackage{booktabs}
\usepackage{color,soul}
\usepackage{epstopdf}
\usepackage{float}
\usepackage{bm}
\usepackage{makecell}
\usepackage{array}
\usepackage{bm}
\usepackage{mathtools}
\usepackage{geometry}
\usepackage[linesnumbered,ruled,commentsnumbered,longend]{algorithm2e}
\usepackage[linesnumbered,ruled,commentsnumbered,longend]{algorithm2e}
\makeatletter

\newcommand{\Rmnum}[1]{\expandafter\@slowromancap\romannumeral #1@}

\makeatother
\makeatletter

\newcommand{\qed}{\nobreak \ifvmode \relax \else
	\ifdim\lastskip<1.5em \hskip-\lastskip
	\hskip1.5em plus0em minus0.5em \fi \nobreak
	\vrule height0.75em width0.5em depth0.25em\fi}

\SetAlgoLongEnd

\ifCLASSINFOpdf
\else
\fi
\begin{document}

\title{\LARGE{Machine Learning-based Beamforming Design for Millimeter Wave IRS Communications with Discrete Phase Shifters}}
\author{Wencai Yan, Gangcan Sun, Wanming Hao,  Zhengyu Zhu, Zheng Chu, and Pei Xiao
	\thanks{W. Yan, G. Sun, W. Hao and Z. Zhu are with the School of Information Engineering, Zhengzhou University, Zhengzhou 450001, China. (E-mail: yanwencai001@163.com, \{iegcsun, iewmhao, iezyzhu\}@zzu.edu.cn)}
	\thanks{Z. Chu and P. Xiao is with the 5G Innovation Centre, Institute of Communication Systems, University of Surrey, Guildford GU2 7XH, U.K. (E-mail: andrew.chuzheng7@gmail.com, p.xiao@surrey.ac.uk).}}
\maketitle
\begin{abstract}
In this paper, we investigate an intelligent reflecting surface (IRS)-assisted millimeter-wave multiple-input single-output downlink wireless communication system. By jointly calculating the active beamforming at the base station and the passive beamforming at the IRS, we aim to minimize the transmit power under the constraint of each user' signal-to-interference-plus-noise ratio. To solve this problem, we propose a low-complexity machine learning-based cross-entropy (CE) algorithm to alternately optimize the active beamforming and the passive beamforming. Specifically, in the alternative iteration process, the zero-forcing (ZF) method and CE algorithm are applied to acquire the active beamforming and the passive beamforming, respectively. The CE algorithm starts with random sampling, by the idea of distribution focusing, namely shifting the distribution towards a desired one by minimizing CE, and a near optimal reflection coefficients with adequately high probability can be obtained. In addition, we extend the original one-bit phase shift at the IRS to the common case with high-resolution phase shift to enhance the effectiveness of the algorithms.
Simulation results verify that the proposed algorithm can obtain a near optimal solution with lower computational complexity.
\end{abstract}


%
\IEEEpeerreviewmaketitle

\section{Introduction}
Millimeter-wave (mmWave) and intelligent reflecting surface (IRS) have recently been considered as two promising technologies for supporting the high-speed data transmission in future wireless communications~\cite{ref1}-\cite{ref3}. Because of the small wavelength of the mmWave, a large antenna array can be easily deployed in a compact form to achieve higher array and multiplexing gains. Moreover, the high directive mmWave beam reduces the multi-user interference. However, the severe path loss of the mmWave signals results in limited coverage~\cite{ref3}. To tackle this problem, the IRS, a passive planar surface composing of massive intelligent reflector units, is developed, and the signal transmission direction can be changed by designing the amplitude/phase of the reflector elements so as to improve the coverage. In addition, the IRS is also energy efficient without requiring any extra radio frequency (RF) chain~\cite{ref1}. As such the mmWave IRS communications will be a promising enabling technology for future wireless networks~\cite{ref6}-\cite{ref10}.

One of the main challenges in the IRS communications is beamforming optimization. Primitively, the perfect reflector surface is considered, namely the continuous phase shifts at the IRS and the active beamforming at the base station (BS) are jointly optimized to minminize the transmit power~\cite{ref7}, and a semidefinite relaxation (SDR)-based alternating optimization method is proposed. Instead of single IRS, the authors in~\cite{ref8} consider a cooperative IRS communication scenario. Based on this, an SDR-based bisection method of cooperative passive beamforming at the IRSs and active beamforming at the BS is developed to maximize the minimum signal-to-interference-plus-noise ratio (SINR). Later, considering the practical hardware limit, only the discrete phase shifts at the IRS are applied. The authors of~\cite{ref4} assume that the phase resolution is infinite to make the targeted problem more tractable. The authors of~\cite{ref5} first solve problem with all discrete optimization variables relaxed to their continuous counterparts and then directly quantize each of the obtained continuous phase shifts to its nearest discrete value.
The authors in~\cite{ref9} propose a successive refinement method to formulate the effective passive beamforming by adjusting the discrete phase shifts. Similarly, the authors of~\cite{ref11} also consider the finite-level phase shifts and propose an exhaustive search algorithm. However, for the above schemes, when the number of the reflective elements becomes larger, the computational complexity is extremely high.

\begin{figure}[htbp]
	\centering
		\label{IRS} 
		\includegraphics[width=6.5cm,height=3.2cm]{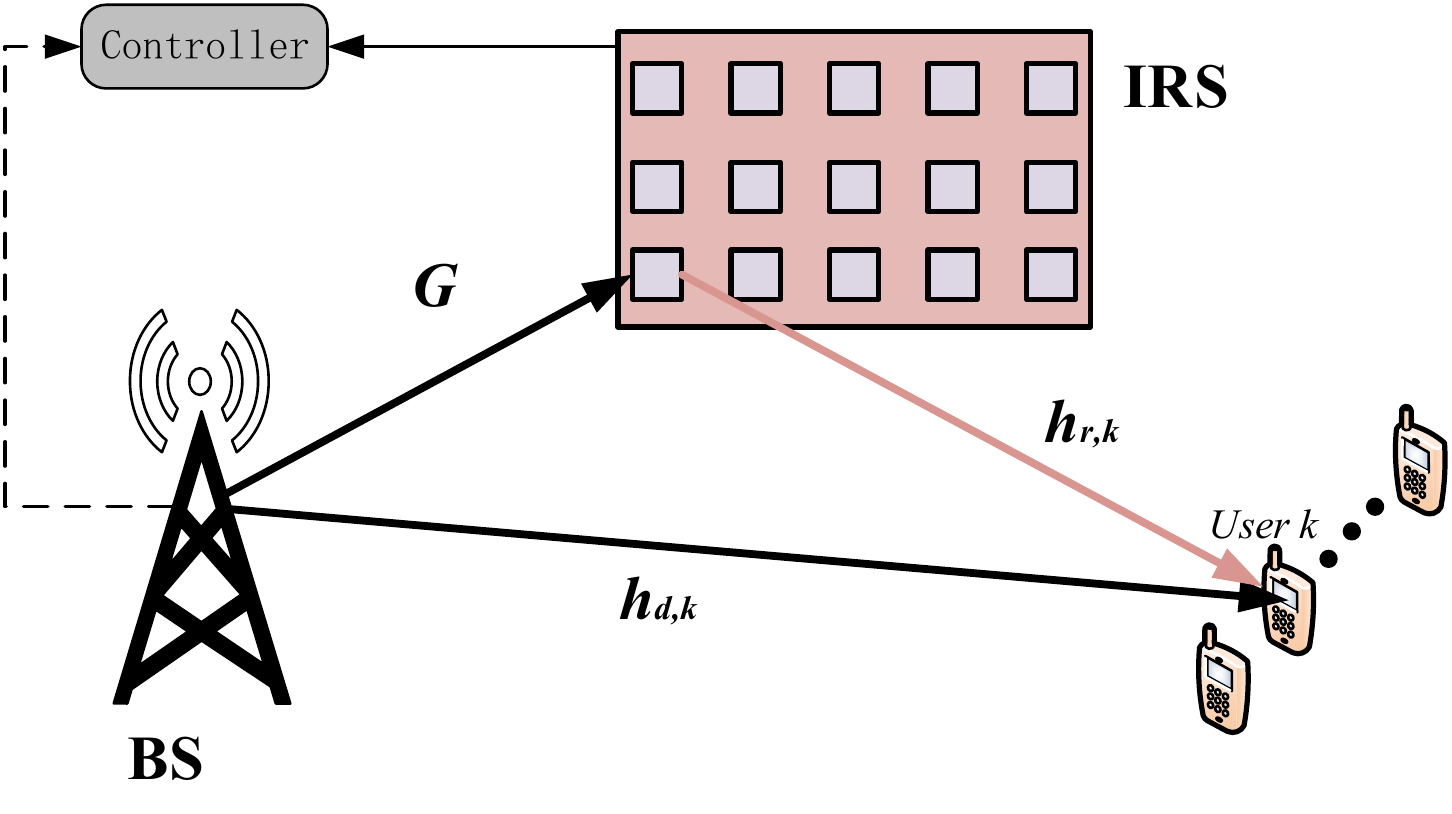}
	\caption{The downlink mmWave IRS communication systems}
\end{figure}

To reduce the computational complexity of optimizing the discrete phase shifts, in this paper, a low-complexity machine learning-based optimization scheme in the mmWave IRS communications is proposed. By jointly optimizing the active beamforming at the BS and passive beamforming at the IRS, we first consider a one-bit phase shift at the IRS, and formulate a problem of minimizing the transmit power under the constraint of user' SINR. Specifically, on the basis of the probability distributions of the reflection coefficients at the IRS, we first randomly generate several groups of IRS reflection coefficients. Then, based on the generated reflection coefficients, the zero-forcing (ZF) method is applied to obtain the active beamforming at the BS. Next, we calculate the transmit power and select the reflection coefficients corresponding to minimum transmit power to update the probability distributions of the reflection coefficients at the IRS by minimizing CE. The above procedure starts with random initialization, but by iteratively minimizing the CE distance, and it becomes a highly effective learning algorithm. The above steps are repeated until convergence, such that near optimal reflection coefficients and active beamforming are obtained. Finally, we extend the CE algorithm to the common case with high-resolution phase shifts. Experimental results verify that the proposed algorithm strikes a good balance between performance and computational complexity.
\section{System Model and Problem Formulation}
\subsection{System Model}
As shown in Fig.~1, we consider a downlink mmWave IRS communication system, which comprises an {\it{M}}-antenna BS, an {\it{N}}-element IRS and  {\it{K}} single-antenna users. Meanwhile, the BS and the IRS are deployed as uniform planer array (UPA). We assume that the CSIs of all links can be obtained using existing channel estimation schemes~\cite{ref7},~\cite{ref15}. The mmWave channels of the BS-IRS link, the IRS-user link and the BS-user link are, respectively denoted by $\mathbf{G} \in \mathbb{C}^{N \times M}$, $\mathbf{h}_{r, k} \in \mathbb{C}^{N \times 1}$ and $\mathbf{h}_{d, k} \in \mathbb{C}^{M \times 1}$. $\mathbf{G}$ can be expressed as~\cite{ref15}
\begin{eqnarray}
  \mathbf{G}=\sqrt{\frac{M N}{L_{G}}} \sum_{l_{1}}^{L_{G}} \alpha_{l_{1}}^{G} \mathbf{b}\left(\vartheta_{l_{1}}^{G_{t}}, \psi_{l_{1}}^{G_{t}}\right) \mathbf{a}\left(\vartheta_{l_{1}}^{G_{r}}, \psi_{l_{1}}^{G_{r}}\right)^{T},
\end{eqnarray}
where $L_{G}$ denotes the paths number, $\alpha_{l_{1}}^{G}$, $\vartheta_{l_{1}}^{G_{t}}\left(\psi_{l_{1}}^{G_{t}}\right)$ and $\vartheta_{l_{1}}^{G_{r}}\left(\psi_{l_{1}}^{G_{r}}\right)$  denote the complex gain, the azimuth (elevation) angle of departure, and the azimuth (elevation) angle of arrival for the $l_{1}$th path, respectively. Similarly, $\mathbf{h}_{r, k}$ and $\mathbf{h}_{d, k}$ are given by
\begin{eqnarray}
  \mathbf{h}_{r, k}=\sqrt{\frac{N}{L_{r, k}}} \sum_{l_{2}=1}^{L_{r, k}} \alpha_{l_{2}}^{r, k} \mathbf{a}\left(\vartheta_{l_{2}}^{r, k}, \psi_{l_{2}}^{r, k}\right),
\end{eqnarray}
\begin{eqnarray}
  \mathbf{h}_{d, k}=\sqrt{\frac{M}{L_{d, k}}} \sum_{l_{3}=1}^{L_{d, k}} \alpha_{l_{3}}^{d, k} \mathbf{b}\left(\vartheta_{l_{3}}^{d, k}, \psi_{l_{3}}^{d, k}\right),
\end{eqnarray}
where $L_{r, k}$ denotes the number of paths between the the IRS and the $k$th user, $L_{d, k}$ represents the number of paths between the BS and the $k$th user, $\mathbf{a}(\vartheta, \psi) \in \mathbb{C}^{N \times 1}$ and $\mathbf{b}(\vartheta, \psi) \in \mathbb{C}^{M \times 1}$ represent array steering vectors at the IRS and the BS, respectively. For the IRS, we use a UPA with $N_1$ elements in horizon and $N_2$ elements in vertical $\left({N=N_1}{N_2}\right)$, and thus
$\mathbf{a}(\vartheta, \psi)$ is expressed~as
\begin{eqnarray}
\mathbf{a}(\vartheta, \psi)=\frac{1}{\sqrt{N}}\left[e^{-j 2 \pi d \sin (\vartheta) \cos (\psi) \mathbf{n}_{1} / \lambda}\right] \otimes\left[e^{-j 2 \pi d \sin (\psi) \mathbf{n}_{2} / \lambda}\right],
\end{eqnarray}
where $\mathbf{n}_{1}=\left[0,1, \cdots, N_{1}-1\right]$ and $\mathbf{n}_{2}=\left[0,1, \cdots, N_{2}-1\right]$, $\lambda$ represents the signal wavelength, $d$ denotes the element spacing and usually set as $d=\lambda / 2$.  $\mathbf{b}(\vartheta, \psi)$ has the similar expression, and we omit it due to the limited space.

The received signal at user $k$ is formulated as
\begin{eqnarray}
y_{k}=\left(\mathbf{h}_{r, k}^{H} \mathbf{\Phi} \mathbf{G}+\mathbf{h}_{d, k}^{H}\right) \sum\nolimits_{j=1}^{K} \mathbf{w}_{j} s_{j}+n_{k}, \quad k \in \left[0,1, \cdots, K\right],
\end{eqnarray}
where $\mathbf{\Phi}=\operatorname{diag}\left(\varphi_{1}, \cdots, \varphi_{N}\right)$ is the $N \times N$ diagonal reflection matrix of the IRS, reflection coefficient $\varphi_{n}$ is the $n$th element of the IRS and defined as $\varphi_{n}= \beta_{n}e^{j \theta_{n}}$, in which $\beta_{n} \in[0,1]$ and $\theta_{n} \in[0, 2 \pi)$  respectively denote the amplitude reflection coefficient and the reflection phase shift, $\mathbf{w}_{k} \in \mathbb{C}^{M \times 1}$ is the active beamforming at the BS for user $k$, $s_{k}$ represents the message intended to be received by user $k$ satisfying $E\left[\left|s_{k}\right|^{2}\right]=1$, $n_{k} \sim \mathcal{C} \mathcal{N}\left(0, \sigma_{k}^{2}\right)$  is the additive zero average white Gaussian noise (AWGN) with variance of $\sigma_{k}^{2}$ at user $k$. For the IRS, we only optimize the phase shifts and set $\beta_{n}=1, n \in \left[0,1, \cdots, N\right]$ for maximizing the reflection efficiency~\cite{ref9}. Considering the practical hardware limit, we assume that $Q$-bit finite-level phase shift is deployed at each element and employ the uniform quantization method to produce $2^{Q}$ discrete phase-shift values. Therefore, the set of discrete phase-shift values $\boldsymbol{\theta}$ is given by
\begin{eqnarray}
\mathcal{S}_{\theta}=\{0, \Delta \theta, \cdots,(2^{Q}-1) \Delta \theta\},
\end{eqnarray}
where $\Delta \theta=2 \pi / 2^{Q}$. The set of discrete reflection coefficient $\boldsymbol{\varphi}$ values is denoted by
\begin{eqnarray}
\mathcal{S}_{\varphi} = \{e^{j 0}, e^{j{\Delta \theta}}, \cdots, e^{j{\left(2^{Q}-1\right)\Delta \theta}}\}.
\end{eqnarray}
The SINR of user $k$ is formulated as
\begin{eqnarray}
\begin{split}
\operatorname{SINR}_{k}&=\frac{\left|\left(\mathbf{h}_{r, k}^{H} \mathbf{\mathbf{\Phi}} \mathbf{G}+\mathbf{h}_{d, k}^{H}\right) \mathbf{w}_{k}\right|^{2}}{\sum_{j \neq k}^{K}\left|\left(\mathbf{h}_{r, k}^{H} \mathbf{\Phi} \mathbf{G}+\mathbf{h}_{d, k}^{H}\right) \mathbf{w}_{j}\right|^{2}+\sigma_{k}^{2}}.\\
\end{split}
\end{eqnarray}
\subsection{Problem Formulation}
We aim to calculate the active beamforming $\mathbf{W}$ at the BS and the reflection coefficients $\boldsymbol{\varphi}$ at the IRS to minimize the transmit power of the BS, where $\boldsymbol{\varphi}=\left[\varphi_{1}, \cdots, \varphi_{N}\right]$ and $\mathbf{W}=\left[\mathbf{w}_{1}, \cdots, \mathbf{w}_{K}\right] \in\mathbb{C}^{M \times K}$. The optimization problem is given by
\begin{subequations}\label{OptA}
\begin{align}
\;\;\;\;\;\;\;\;\mathrm{P1}:&\min _{\mathbf{W}, \boldsymbol{\varphi}} \sum_{k=1}^{K}\left\|\mathbf{w}_{k}\right\|^{2}\label{OptA0}\\
{\rm{s.t.}}\;\;&\frac{\left|\left(\mathbf{h}_{r, k}^{H} \mathbf{\Phi} \mathbf{G}+\mathbf{h}_{d, k}^{H}\right) \mathbf{w}_{k}\right|^{2}}{\sum_{j \neq k}^{K}\left|\left(\mathbf{h}_{r, k}^{H} \boldsymbol{\mathbf{\Phi}} \mathbf{G}+\mathbf{h}_{d, k}^{H}\right) \mathbf{w}_{j}\right|^{2}+\sigma_{k}^{2}} \geq \gamma_{k},\forall k,\label{OptA1}\\
&\varphi_{n} \in \mathcal{S}_{\varphi}, \quad n \in \left[0,1, \cdots, N\right],\label{OptA2}
\end{align}
\end{subequations}
where $\gamma_{k}$ is the SINR requirement of user $k$. Note that the optimal solution to P1 is intractable due to the non-convex constraints. The non-convexity is caused by the coupled variables (i.e., $\mathbf{W}$ and $\boldsymbol{\varphi}$) and the discrete values of the reflection coefficient $\boldsymbol{\varphi}$. To solve it, one generally adopted method is iterative optimization algorithm. Specifically, for any given reflection coefficient $\boldsymbol{\varphi}$, the active beamforming $\mathbf{W}$ is obtained by employing the minimum mean squared error (MMSE) or the suboptimal ZF-based method with low computational complexity. Then, $\mathbf{W}$ and $\boldsymbol{\varphi}$ are updated alternatively until convergence. However, existing algorithms for optimizing reflection coefficient $\boldsymbol{\varphi}$ are usually based on the exhaustive search or successive refinement~\cite{ref9}, which results in higher computational complexity. To work around this issue, we will develop a machine learning-based CE algorithm with low complexity to obtain the reflection coefficient $\boldsymbol{\varphi}$.
\section{Proposed Solution}
In this section, we propose a machine learning-based CE algorithm and utilize an alternatively iterative method to solve problem $\mathrm{P1}$. We first randomly generate several reflection coefficients $\boldsymbol{\varphi}$ based on the probability distributions. Based on the generated reflection coefficients $\boldsymbol{\varphi}$, we solve the active beamforming $\mathbf{W}$ by applying the ZF scheme for obtaining a tradeoff between performance and complexity. Next, we update the probability distributions of the reflection coefficient at the IRS to regenerate the reflection coefficients $\boldsymbol{\varphi}$. Repeating above steps, $\mathbf{W}$ and $\boldsymbol{\varphi}$ are optimized alternately until convergence.

We first introduce the machine learning-based CE algorithm, and the aim is to obtain the (approximate) optimal performance for a learning task~\cite{ref12}-\cite{ref14}. It comprises two phases: 1) Generating several random data on the basis of a specified probability distribution; 2) Based on certain criteria (e.g., minimum transmit power in this work), selecting several best data as ``elite" to update the probability distribution parameters by minimizing the CE. Repeating such procedure, the probability distribution will be refined to generate a solution close to the optimal one with a sufficiently high probability. Next, the specific steps of the proposed algorithm as follows:

When $Q = 1$,  the phase shift $\theta_{n} \in \mathcal{S}_{\theta} = \{0,\pi\}$, the reflection coefficient $\varphi_{n} \in \mathcal{S}_{\varphi} = \{-1,+1\}$. First, we define the probability parameter $\mathbf{p}=\left[{p}_{1}, \cdots, {p}_{N}\right]$ as a $1 \times N$ vector, where $0 \leq {p}_{n} \leq 1$ denotes the probability of reflection coefficient $\varphi_{n} = 1$. Because there is no priori information, we initialize the parameter $\mathbf{p}^{(0)}=\frac{\mathbf{1}}{\mathbf{2}} \times \mathbf{1}_{1 \times N}$, where $\mathbf{1}$ is a vector with all-one elements. In other words, in the initialization phase, we assume that all the reflection coefficients belong to $\{-1,+1\}$  with equal probability. At the $i$th iteration, we generate $S$ candidate reflection coefficients $\{\boldsymbol{\varphi}^{[s]}\}_{s=1}^S$ according to the probability distribution $\mathcal{T}\left(\mathcal{S}_{\varphi} ; \mathbf{p}^{(i)}\right)$. Under any given reflection coefficients $\boldsymbol{\varphi}^{[s]}$, the transmit power minimization problem $\mathrm{P1}$ can be formulated as
\begin{subequations}\label{OptB}
\begin{align}
\;\;\;\;\;\;\;\;\mathrm{P2}:&\min _\mathbf{W} \sum\nolimits_{k=1}^{K}\left\|\mathbf{w}_{k}\right\|^{2}\label{OptB0}\\
{\rm{s.t.}}\;\;&\frac{\left|\left(\mathbf{h}_{r, k}^{H} \mathbf{\Phi} \mathbf{G}+\mathbf{h}_{d, k}^{H}\right) \mathbf{w}_{k}\right|^{2}}{\sum_{j \neq k}^{K}\left|\left(\mathbf{h}_{r, k}^{H} \boldsymbol{\mathbf{\Phi}} \mathbf{G}+\mathbf{h}_{d, k}^{H}\right) \mathbf{w}_{j}\right|^{2}+\sigma_{k}^{2}} \geq \gamma_{k},\forall k.\label{OptB1}
\end{align}
\end{subequations}
We apply the low-complexity ZF technique to acquire the active beamforming $\mathbf{W}$, which can be computed as
\begin{eqnarray}
\mathbf{W}=\mathbf{H}^{H}\left(\mathbf{H} \mathbf{H}^{H}\right)^{-1} \mathbf{U}^{\frac{1}{2}},
\end{eqnarray}
where $\mathbf{H}=\mathbf{H}_{r}^{H} \mathbf{\Phi} \mathbf{G}+\mathbf{H}_{d}^{H}$, $\mathbf{H}_{r}^{H}=\left[\mathbf{h}_{r, 1}, \cdots, \mathbf{h}_{r, K}\right]^{H}$ and $\mathbf{H}_{d}^{H}=\left[\mathbf{h}_{d, 1}, \cdots, \mathbf{h}_{d, K}\right]^{H}$, $\mathbf{U}=\operatorname{diag}\left(u_{1}, \cdots, u_{K}\right)$ is the power allocation matrix.
\begin{algorithm}
\caption{The proposed machine learning-based CE algorithm}
\hspace*{0.02in} {\bf Input:}
Channel matrix $\mathbf{G}$, $\mathbf{h}_{r, k}$, $\mathbf{h}_{d,k}$; Number of candidates and elites $S$, $S_\text{elite}$; Number of iterations $I$.\\
\hspace*{0.02in} {\bf Initialization:}
$i=0$, $\mathbf{p}^{(0)}=\frac{\mathbf{1}}{\mathbf{2}} \times \mathbf{1}_{N \times 1}.$

\hspace*{0.02in} {\bf for $i=0$ to $I$ do }\\
Generate $S$ random candidate reflection coefficients $\{\boldsymbol{\varphi}^{[s]}\}_{s=1}^S$ according to the $\mathcal{T}\left(\mathcal{S}_{\varphi} ; \mathbf{p}^{(i)}\right)$;\\
Obtain the active beamforming $\mathbf{W}$ based on (11);\\
Calculate the $S$ corresponding transmit power $\{\mathcal{P}\left( \boldsymbol{\varphi}^{[s]} \right)\}_{s=1}^S$ based on (12);\\
Sort $\{\mathcal{P}\left( \boldsymbol{\varphi}^{[s]} \right)\}_{s=1}^S$ in a descend order and select $S_\text{elite}$ reflection coefficients with the lowest transmit power as elite samples;\\
Update $\mathbf{p}^{(i+1)}$ based on (17);\\
\hspace*{0.02in} {\bf end}

\hspace*{0.02in} {\bf Output:}
$\mathbf{W}$, $\boldsymbol{\varphi}$.
\end{algorithm}
According to~\cite{ref9}, the constraint (10b) can be reformed as ${u_{k}}/{\sigma_{k}^{2}} \geq \gamma_{k},\forall k$. To obtain the optimal solution,  we have $u_{k}=\sigma_{k}^{2} \gamma_{k}, \forall k$. The transmit power of BS can be computed as
\begin{eqnarray}
\begin{split}
\sum\nolimits_{k=1}^{K}\left\|\mathbf{w}_{k}\right\|^{2}=\operatorname{tr}\left(\mathbf{U}^{\frac{1}{2}}\left(\mathbf{H} \mathbf{H}^{H}\right)^{-1} \mathbf{U}^{\frac{1}{2}}\right)=\operatorname{tr}\left(\mathbf{U}\left(\mathbf{H} \mathbf{H}^{H}\right)^{-1}\right)=\mathcal{P}(\boldsymbol{\varphi}).
\end{split}
\end{eqnarray}
Therefore, the problem $\mathrm{P2}$ can be transformed to
\begin{subequations}\label{OptC}
\begin{align}
\;\;\;\;\;\;\;\;\mathrm{P3}:&\min _{\boldsymbol{\varphi}} \mathcal{P}(\boldsymbol{\varphi})\label{OptC0}\\
{\rm{s.t.}}\;\;&\boldsymbol{\varphi} \in \{\boldsymbol{\varphi}^{[1]}, \cdots, \boldsymbol{\varphi}^{[S]}\}.\label{OptC1}
\end{align}
\end{subequations}
Then, we compute $S$ transmit power $\{\mathcal{P}\left( \boldsymbol{\varphi}^{[s]} \right)\}_{s=1}^S$ corresponding to the candidate reflection coefficients $\{\boldsymbol{\varphi}^{[s]}\}_{s=1}^S$ based on (12) and sort $\{\mathcal{P}\left( \boldsymbol{\varphi}^{[s]} \right)\}_{s=1}^S$ in a descend order. After that, we choose $S_{elite}$ reflection coefficients with the lowest transmit power as elite samples to update $\mathbf{p}^{(i+1)}$ by minimizing CE
\begin{eqnarray}
\mathbf{p}^{(i+1)}=\arg \max _{\mathbf{p}^{(i)}} \frac{1}{S} \sum_{s=1}^{S_{\text {elite }}} \ln \mathcal{T}\left(\mathcal{S}_{\varphi}^{[s]} ; \mathbf{p}^{(i)}\right).
\end{eqnarray}
Note that the $n$th element $\varphi_{n}^{[s]}$ of $\boldsymbol{\varphi}^{[s]}$ complies with Bernoulli distribution. We set the probability of $\varphi_{n}^{[s]} = +1$ as $p_{n}^{(i)}$ and the probability of $\varphi_{n}^{[s]} = -1$ as $1-p_{n}^{(i)}$. Thus  the probability distribution $\mathcal{T}\left(\mathcal{S}_{\varphi}^{[s]} ; \mathbf{p}^{(i)}\right)$ is given by
\begin{eqnarray}
\mathcal{T}\left(\mathcal{S}_{\varphi}^{[s]} ; \mathbf{p}^{(i)}\right)=\prod_{n=1}^{N}\left(p_{n}^{(i)}\right)^{\frac{1}{2}\left(1+\varphi_{n}^{[s]}\right)}\left(1-p_{n}^{(i)}\right)^{\frac{1}{2}\left(1-\varphi_{n}^{[s]}\right)}.
\end{eqnarray}
By substituting (15) into (14), the first derivative of (14) with respect to $p_{n}^{(i)}$ can be calculated as
\begin{eqnarray}
\frac{1}{S} \sum_{s=1}^{S_{\text {elite }}} \left(\frac{1+ \varphi_{n}^{[s]}}{2 p_{n}^{(i)}}-\frac{1- \varphi_{n}^{[s]}}{2\left(1-p_{n}^{(i)}\right)}\right).
\end{eqnarray}
Setting (16) as zero, $p_{n}^{(i+1)}$ is updated by
\begin{eqnarray}
p_{n}^{(i+1)}=\frac{\sum_{s=1}^{S_{\text {elite }}}\left(\varphi_{n}^{[s]}+1\right)}{2 S_{\text {elite }}}.
\end{eqnarray}
The new probability distributions is employed to regenerate $S$ candidate reflection coefficients. The above steps are repeated until the number of iterations $i=I$, where $I$ is the number of iterations required for convergence. Finally, near optimal active beamforming $\mathbf{W}$ and reflection coefficients $\boldsymbol{\varphi}$ with minimum transmit power are obtained. We summarize the above steps in Algorithm $\boldsymbol{1}$. Several best data can be selected as ``elite" to update the probability distribution parameters at each iteration, and it guarantees that the solution tends to be optimal.  Moreover, it has been shown in~\cite{ref13} that the CE method terminates with probability 1 in a finite number of iterations.

The computational complexity of Algorithm $\boldsymbol{1}$ is mainly due to: 1) In step $5$, we obtain the active beamforming $\mathbf{W}$ based on (11) with the complexity $\mathcal{O}\left(N S K^{2}\right)$. 2) In step $6$, we calculate $S$ corresponding transmit power $\{\mathcal{P}\left( \boldsymbol{\varphi}^{[s]} \right)\}_{s=1}^S$ according to (12), which involves the complexity $\mathcal{O}\left(S\right)$. 3) The final part is step $8$, we update $\mathbf{p}^{(i+1)}$ on the basis of (17), thus the complexity of this procedure is $\mathcal{O}\left(N S_{elite}\right)$. After $I$ iterations, the total computational complexity of Algorithm $\boldsymbol{1}$ is $\mathcal{O}\left(I N S K^{2}\right)$. It can be observed from simulations that a good performance can be obtained even for a small $I$ and $S$.

Next, we investigate a more common scenario where $Q \textgreater 1$. In this case, the phase shift $\theta_{n} \in \mathcal{S}_{\theta} = \{0, \Delta \theta, \cdots,(2^{Q}-1) \Delta \theta\}$, the reflection coefficient $\varphi_{n} \in \mathcal{S}_{\varphi} = \{e^{j 0}, e^{j{\Delta \theta}}, \cdots, e^{j{\left(2^{Q}-1\right)\Delta \theta}}\}$, we can generate a random sample $\left\{\varphi_{n}\right\}_{n=1}^{N}$, $\varphi_{n}$ is independently drawn from the set $\mathcal{S}_{\varphi}$ and follows discrete distribution $\left\{p_{q, n}^{(i)}\right\}_{q=1}^{2^{Q}}$, where $\left\{p_{q, n}^{(i)}\right\}$ is the probability of the $q$th element in set $\mathcal{S}_{\varphi}$ being selected as $\varphi_{n}$. Here, $\left\{p_{q, n}^{(i)}\right\}$ satisfies $\sum_{q=1}^{2^{Q}} p_{q, n}^{(i)}=1$.
We still assume that all the reflection elements belong to the set $\mathcal{S}_{\varphi}$  with equal probability in the initialization phase.

The next step involves employing the selected elite samples to update $p_{q,n}^{(i+1)}$ by minimizing CE
\begin{eqnarray}
p_{q,n}^{(i+1)}=\arg \max _{p_{q,n}^{(i)}} \frac{1}{S} \sum_{s=1}^{S_{\text {elite }}} \ln \mathcal{T}\left(\mathcal{S}_{\varphi}^{[s]} ; p_{q,n}^{(i)}\right),
\end{eqnarray}
where $\mathcal{T}\left(\mathcal{S}_{\varphi}^{[s]} ; p_{q,n}^{(i)}\right)$ is calculated as
\begin{eqnarray}
\mathcal{T}\left(\mathcal{S}_{\varphi}^{[s]} ; p_{q,n}^{(i)}\right)=\prod_{n=1}^{N} \sum_{q=1}^{2^{Q}} p_{q, n}^{(i)} \mathbbm{1}_{\left\{\varphi_{n}^{[s]} = \mathcal{S}_{\varphi,q}\right\}},
\end{eqnarray}
where $\mathcal{S}_{\varphi,q}$ denotes the $q$th element in set $\mathcal{S}_{\varphi}$, and the function $\mathbbm{1}_{\left\{\bullet \right\}}= 1$ when the constraint ${\left\{\bullet \right\}}$ is satisfied, otherwise $\mathbbm{1}_{\left\{\bullet \right\}}= 0$. Furthermore, to satisfy the constraint of $\sum_{q=1}^{2^{Q}} p_{q, n}^{(i)}=1$, we introduce the Lagrange multiplier $\left\{\mathcal{L}_{n}\right\}_{n=1}^{N}$ into (18) as
\begin{eqnarray}
\begin{split}
p_{q,n}^{(i+1)}&=\arg \max _{p_{q,n}^{(i)}} \frac{1}{S} \sum_{s=1}^{S_{\text {elite }}} \ln \mathcal{T}\left(\mathcal{S}_{\varphi}^{[s]} ; p_{q,n}^{(i)}\right)\\
&+ \sum_{n=1}^{N} \mathcal{L}_{n}\left(\sum_{q=1}^{2^{Q}}\left(p_{q, n}^{(i)}-1\right)\right).
\end{split}
\end{eqnarray}
Taking the first derivative of (20) with respect to $p_{q,n}^{(i)}$ and setting the result as zero, yield
\begin{eqnarray}
\frac{1}{S} \sum_{s=1}^{S_{\text {elite }}} \mathbbm{1}_{\left\{\varphi_{n}^{[s]} = \mathcal{S}_{\varphi,q}\right\}}+\mathcal{L}_{n} p_{q, n}^{(i)}=0.
\end{eqnarray}
By adding all of (21) for $q = 1,2, \cdots, 2^{Q}$, we have $\mathcal{L}_{n} = - \frac{1}{S} S_{\text {elite }}$.
Then, by substituting (21) into (20), we obtain
\begin{eqnarray}
p_{q, n}^{(i+1)} = \frac{\sum_{s=1}^{S_{\text {elite }}} \mathbbm{1}_{\left\{\varphi_{n}^{[s]} = \mathcal{S}_{\varphi,q}\right\}}}{S_{\text {elite }}}.
\end{eqnarray}

\begin{figure*}[t]
	\centering
	\subfigure[]{
		\label{Iteration} 
		\includegraphics[width=6.5cm,height=4.15cm]{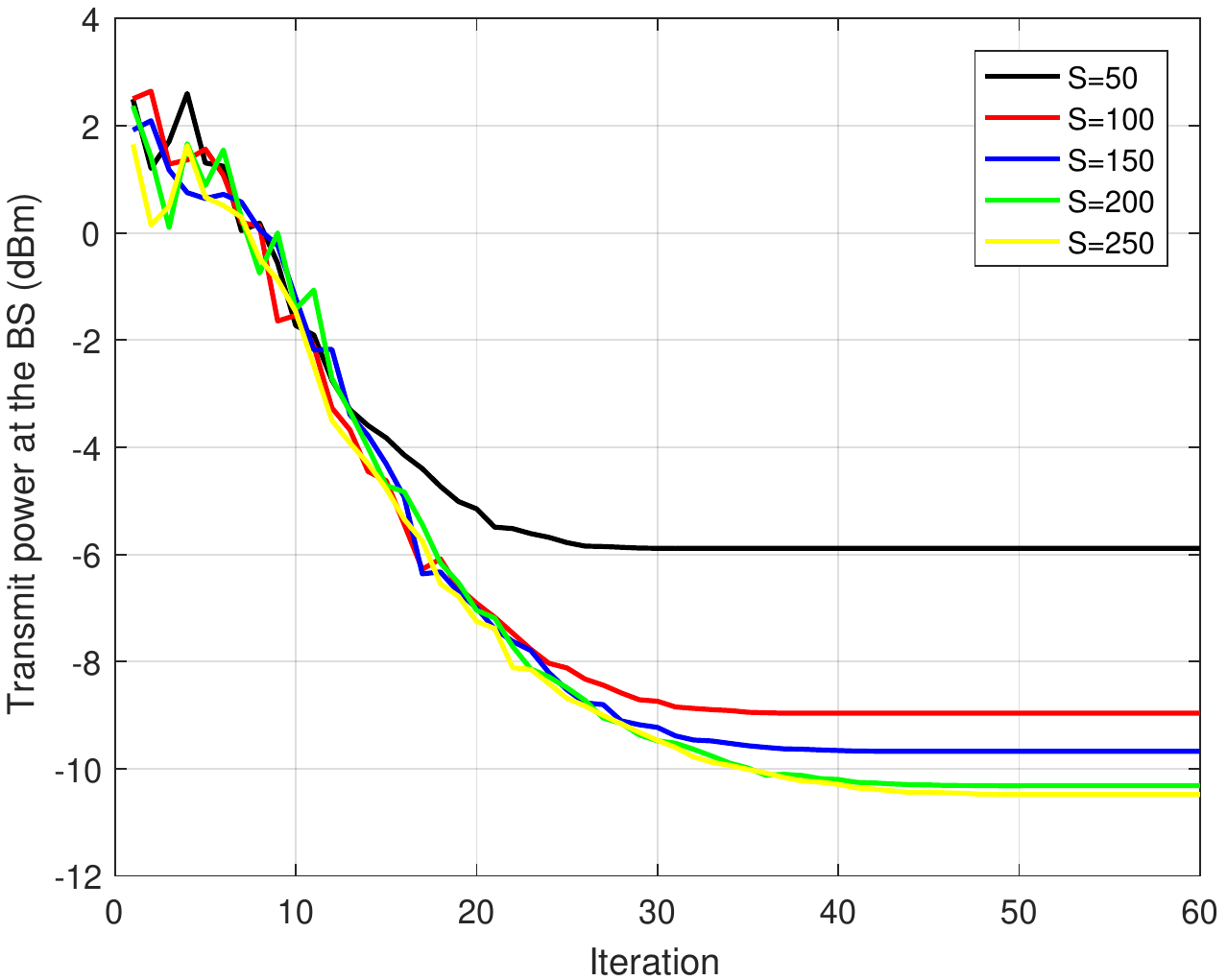}}
	\subfigure[]{
		\label{complexity} 
		\includegraphics[width=6.5cm,height=4.15cm]{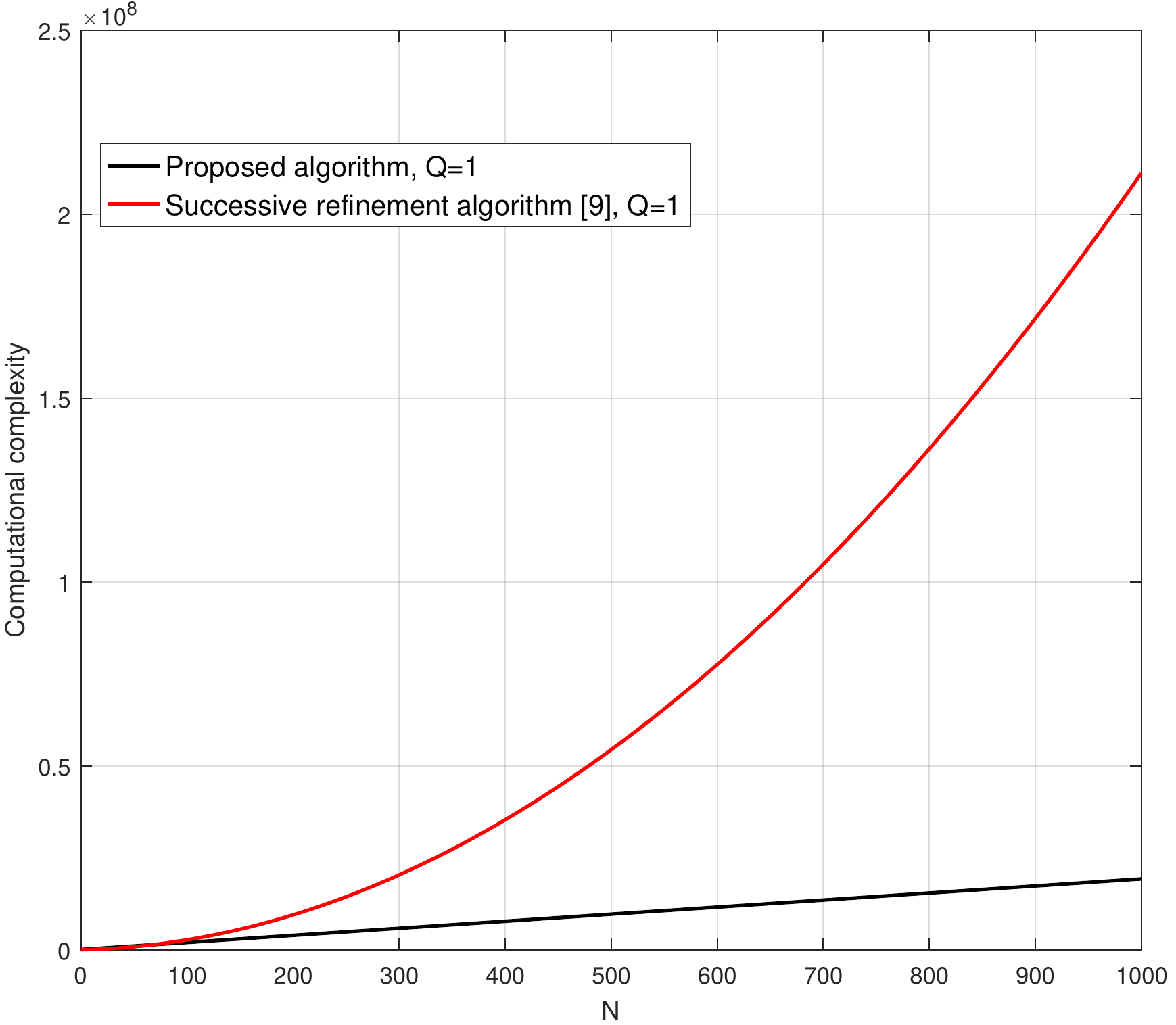}}
	\subfigure[]{
		\label{exhaustive} 
		\includegraphics[width=6.5cm,height=4.2cm]{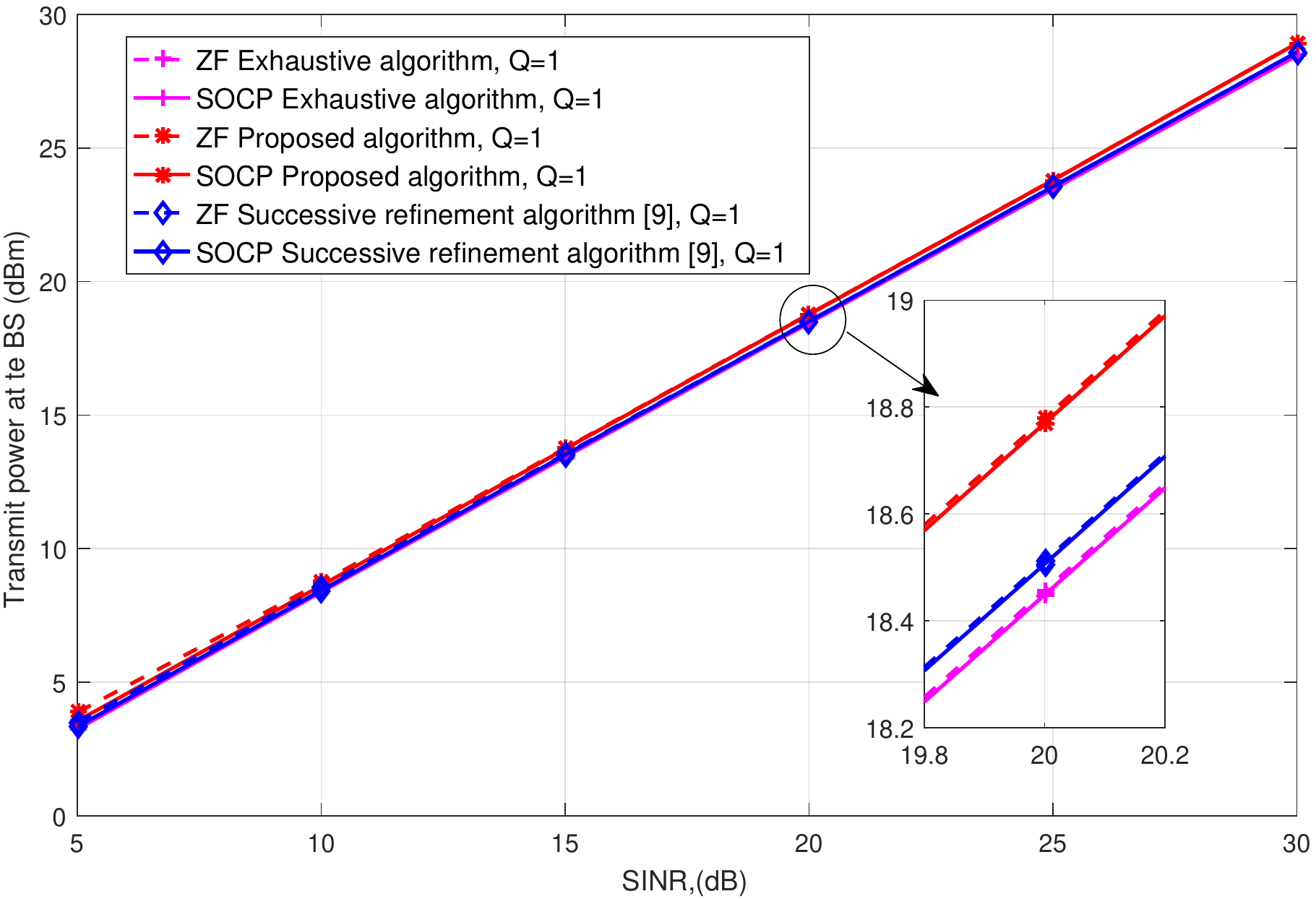}}
    \subfigure[]{
		\label{SINR} 
		\includegraphics[width=6.5cm,height=4.2cm]{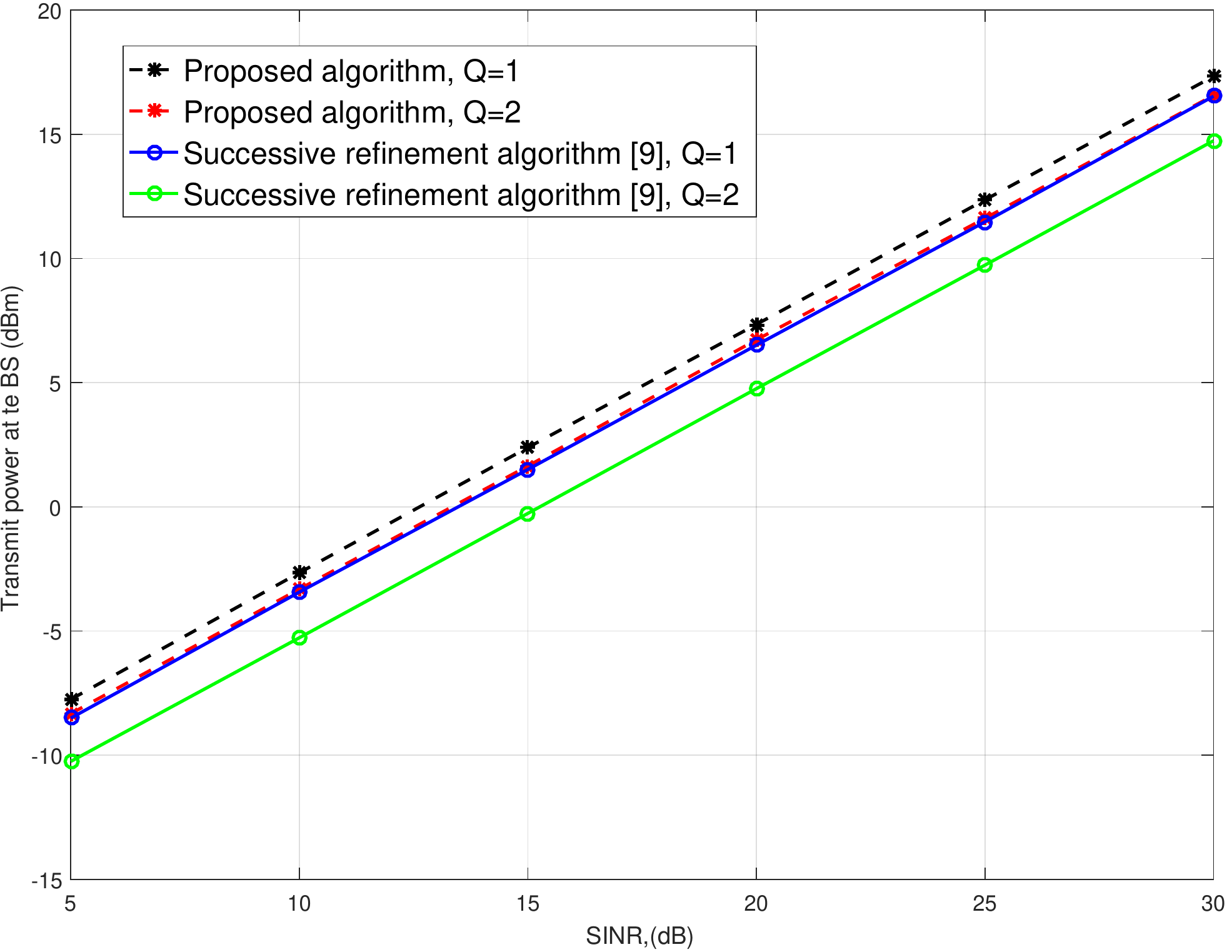}}
	\caption{(a) Transmit power versus iterations $I$. (b) Computational complexity comparison versus the number of IRS elements. (c) Transmit power versus SINR, $M=4$, $K = 2$, $N=8$. (d) Transmit power versus SINR, $M=64$, $K = 4$, $N=625$.}
\end{figure*}
\section{Numerical Results}
In this section, we evaluate the performance of the proposed machine learning-based CE algorithm. In the simulation, we set $L_{G}=4$, $L_{r, k}=5$ and $L_{d, k}=3$ for $\forall k$. The distance-dependent channel path loss of BS-IRS link, IRS-user link and BS-user link are $10^{-3} d_{B R}^{-2.2}$, $10^{-3} d_{R U}^{-2.8}$, $10^{-3} d_{B U}^{-3.5}$, respectively. $d_{B R}$, $d_{R U}$, $d_{B U}$ respectively denote the distance of BS-IRS link, IRS-user link and BS-user link. We deploy $d_{B R}=50m$, $d_{R U}=2m$, $d_{B U}=60m$ for $\forall k$. We set that the noise power to $\sigma_{k}^{2} = -90$dBm for $\forall k$ and all users have the same SINR requirement. Besides, we deploy UPA at both the BS and IRS. The antenna spacing is set as $d=\lambda / 2$. The carrier frequency is 30 GHz.

First, we evaluate the convergence of the proposed algorithm. Fig.~2(a) shows the BS transmit power versus the iteration number under different number of candidates with $Q = 1$, when $S_{elite}/ S=0.2$, $M=64$, $N = 625$, $K = 4$, and $\gamma = 20$dB. One can observe that the transmit power first decreases and then tends to be stabilized upon a number of iterations, which shows the effectiveness of the proposed algorithm. Meanwhile, it can be seen that a larger S can obtain a lower transmit power. However, it is sufficient to obtain a near optimal performance when $S=200$.

Fig.~2(b) compares the computational complexity between our proposed algorithm and the successive refinement algorithm~\cite{ref9}, where we adopt $S=200$, $S_{elite}=40$, $M=64$, $K = 4$. According to~\cite{ref9}, the computational complexity of the successive refinement algorithm is $\mathcal{O}\left(I_{iter}2^Q\left(K^{3}+K^{2} M+K M N\right)\right)$, where $I_{iter}$ is the number of iterations and $I_{iter}=10 \times N$ according to our simulations, which means that the number of iterations to reach convergence has close relation to the number of the IRS elements. From Fig.~2(b), we can observe that when the number of the IRS elements is large, the computational complexity of the proposed algorithm is much lower than that of the successive refinement algorithm. The main reason is that the successive refinement algorithm needs to optimize all the elements of IRS one by one. In addition, the computational complexity of our algorithm is affected by Q, but the impact is slight compared with the successive refinement algorithm~\cite{ref9}.

Fig.~2(c) shows the transmit power versus SINR for different algorithms. The proposed algorithm is compared with the following schemes: 1) the successive refinement algorithm~\cite{ref9}; 2) the optimal exhaustive search algorithm. In addition, we compared the performance between the second-order cone program (SOCP) and the ZF method for solving the active beamforming.
 Due to the high complexity of the optimal exhaustive search algorithm, we set $M=4$, $K = 2$, $N=8$, $S=10$, $S_{elite}=2$. One can observe that the proposed algorithm can obtain close to optimal solution with very low computational complexity.

Fig.~2(d) plots the transmit power versus SINR with $Q = 1$, $Q = 2$ for our proposed algorithm and the successive refinement algorithm~\cite{ref9}. We set $S=200$, $S_{elite}=40$, $M=64$, $N=625$, $K = 4$, and $d_{R U}=10m$. As shown in Fig.~1(d), the transmit power is higher under our proposed algorithm than that under the successive refinement algorithm. However, from Fig.~2(b), when $N=625$, we can see that the computational complexity of the successive refinement algorithm is much higher than the proposed algorithm. This demonstrates that our proposed algorithm strikes a good balance between performance and computational complexity. Moreover, the proposed algorithm can achieve a near optimal performance with lower complexity.
\section{Conclusion}
In this paper, under a downlink mmWave IRS communication system, we proposed a low-complexity machine learning-based CE algorithm and utilized an iterative method to alternately optimize the active beamforming at the BS and the passive beamforming at the IRS, and the target is to minimize the transmit power of the BS. We first derived the  probability update expression in the CE algorithm when deploying one-bit phase shift at the IRS and then extended it to high-resolution phase shift setup. The analysis of complexity and simulation results demonstrated that the proposed algorithm can obtain a near optimal performance with a low computational complexity.

\end{document}